\documentclass{emulateapj}

\usepackage{apjfonts}

\newcommand{\msun}{M_\odot}

\catcode`\@=11
\newcommand{\gapprox}{\mathrel{\mathpalette\@versim>}}
\newcommand{\lapprox}{\mathrel{\mathpalette\@versim<}}
\newcommand{\propapprox}{\mathrel{\mathpalette\@versim\propto}}
\newcommand{\@versim}[2]
  {\lower3.1truept\vbox{\baselineskip0pt\lineskip0.5truept
\ialign{$\m@th#1\hfil##\hfil$\crcr#2\crcr\sim\crcr}}}
\catcode`\@=12
\shorttitle{DUST DESTRUCTION IN LMC Core-Collapse SNRS}
\shortauthors{WILLIAMS ET AL.}

\begin{document}

usl\title{Dust Destruction in Fast Shocks 
of Core-Collapse Supernova Remnants in the Large Magellanic Cloud}

\author{Brian J. Williams,\altaffilmark{1}
Kazimierz J. Borkowski,\altaffilmark{1}
Stephen P. Reynolds,\altaffilmark{1}
William P. Blair,\altaffilmark{3}
Parviz Ghavamian,\altaffilmark{3}
Sean P. Hendrick,\altaffilmark{5}
Knox S. Long,\altaffilmark{7}
Sean Points,\altaffilmark{6}
John C. Raymond,\altaffilmark{2}
Ravi Sankrit,\altaffilmark{3,8}
R. Chris Smith,\altaffilmark{6}
\&P. Frank Winkler\altaffilmark{4}
}

\altaffiltext{1}{Physics Dept., North Carolina State U.,
    Raleigh, NC 27695-8202; bjwilli2@ncsu.edu.}
\altaffiltext{2}{Harvard-Smithsonian Center for Astrophysics, 60 Garden
    Street, Cambridge, MA 02138.}
\altaffiltext{3}{Dept. of Physics and Astronomy, Johns Hopkins, 
    3400 N. Charles St., Baltimore, MD 21218-2686.}
\altaffiltext{4}{Dept. of Physics, Middlebury College, Middlebury, VT 
    05753.}
\altaffiltext{5}{Physics Dept., Millersville U., PO Box 1002, Millersville, 
    PA 17551.}
\altaffiltext{6}{NOAO/CTIO, Cailla 603, La Serena, Chile.}
\altaffiltext{7}{STScI, 3700 San Martin Dr., Baltimore, MD 21218.}
\altaffiltext{8}{Current address: UC Berkeley, Space Sciences Laboratory, 7 Gauss Way, Berkeley, CA 94720.}

\begin{abstract}

We report observations with the MIPS instrument aboard the
{\it Spitzer Space Telescope} (SST) of four
supernova remnants (SNRs) believed to be the result of core-collapse
SNe: N132D (0525--69.6), N49B (0525--66.0), N23 (0506--68.0), and
0453--68.5. All four of these SNRs were detected
in whole at 24 $\mu$m and in part at 70 $\mu$m. Comparisons with
{\it Chandra} broadband X-ray images show an
association of infrared (IR) emission with the blast wave. 
We attribute the observed IR emission to dust that has been collisionally heated
by electrons and ions in the hot, X-ray emitting plasma, with grain size
distributions appropriate for the LMC and the destruction of small
grains via sputtering by ions. As with our earlier analysis of Type
Ia SNRs, models can reproduce observed 70/24 $\mu$m flux ratios only
if effects from sputtering are included, destroying small grains. We
calculate the mass of dust swept up by the blast wave in these
remnants, and we derive a dust-to-gas mass ratio of several times less
than the often assumed value of 0.25\% for the LMC. We believe that one
explanation for this discrepancy could be porous (fluffy) dust grains.

\end{abstract}


\section{Introduction}
\label{intro}

Dust plays an important role in all stages of galaxy evolution. The
life-cycle of dust grains and the amount and relative abundances
present in the interstellar medium (ISM) are determined by the
balance between dust formation, grain modification, and dust
destruction \citep{draine03}. Dust destruction is known to
occur in both fast and slow shocks in SNRs \citep{jones04}. We focus
here on dust destruction via sputtering by high energy ions in fast
(non-radiative) shocks in SNRs.  

SNRs make excellent probes of the dust content of the diffuse ISM in
galaxies, since their shock waves create X-ray plasmas that heat dust 
in their vicinities. Ambient dust heated by starlight may escape detection,
but shock-heated dust radiates strongly in {\it Spitzer} wavelengths, so
SNR studies probe different conditions than studies of diffuse ISM dust emission. Modeling
of X-ray emission provides the basis for understanding dust emission.
Inferences of dust content and properties from SNR
observations are thus complementary to UV-absorption studies, which
rely on the fortuitous locations of background UV-bright stars \citep{jenkins84}. 
The combination of these types of investigation may lead to significant 
advances in our understanding of dust properties, with potential repercussions 
for gas-phase abundance determinations, theories of chemical evolution of galaxies,
and dust-catalyzed cosmochemistry.

To examine the nature of dust heating and destruction in the ISM, we
conducted an imaging survey with the {\it SST} of
39 SNRs in the Magellanic Clouds. In a previous paper 
(Borkowski et al.~2006; Paper I) we analyzed IR emission
from SNRs resulting from Type Ia SNe. We found that the dust-to-gas mass 
ratio was lower than the expected value of 0.25\% (Weingartner \& Draine 2001,
hereafter WD), for the surrounding ISM, a result that
 we attributed to Type Ia SNe exploding in lower density media. However, 
\citet{tappe06} found a low dust content in N132D, a remnant of a 
core-collapse SN which occured in a dense ISM area, and suggested a high 
dust destruction efficiency in this SNR. A lower than expected dust content
was also found in the circumstellar medium surrounding SN 1987A 
\citep{bouchet06}. In order to investigate this 
apparent dust deficit, we present here an analysis of four core-collapse 
SNRs, including N132D.

We base our inference of the core-collapse nature of our objects on
the presence of a pulsar-wind nebula in 0453--68.5 \citep{gaensler03}, 
a central compact object in N23 \citep{hayato06, hughes06},
the O-rich classification of N132D \citep{lasker78}, and the presence
of a large mass of Mg in N49B \citep{park03}.

All four of these remnants were detected in both the 24 and 70 $\mu$m 
bands of the Multiband Imaging Photometer for {\it Spitzer} (MIPS).
We find a similar result to that found for the Type Ia remnants in Paper I:
the dust-to-gas ratios we infer are lower by a factor of $\sim 4$ than what is expected for the 
LMC. In section~\ref{concls}, we discuss possible reasons for this 
apparent dust deficiency.

\section{Observations and Data Reduction}
\label{obs}

All four objects were observed with the 24 and 70 $\mu$m MIPS arrays. At 24 $\mu$m, we
mapped each remnant in our survey and the surrounding background with
14 frames of 30.93 seconds each, for a total exposure time of 433
s. At 70 $\mu$m, we observed a total of 545 s in 52 frames for all but
N132D, which we observed for 986 s in 94 frames. MIPS images were
processed from Basic Calibrated Data (BCD) to
Post-BCD (PBCD) at the {\it Spitzer} Science Center (SSC) by version
13.2 of the PBCD pipeline. For the 70 $\mu$m images, we used the
contributed software package GeRT to reprocess the raw telescope
images into BCD images, then reprocessed the BCD images using the SSC
software package MOPEX. GeRT was useful in removing some of the
artifacts, such as vertical stripes, from the 70 $\mu$m data.

Figure ~1 shows the 24 and 70 $\mu$m images as well as X-ray images
from {\it Chandra} archival data and optical images from the Magellanic
Cloud Emission-Line Survey (MCELS; Smith et al.~2005). Because line emission 
from low-ionization gases should be a significant contributor only in 
slower, radiative shocks, we conclude that we are seeing thermal IR 
emission primarily resulting from dust. Recent spectroscopic observations of 
N132D with {\it Spitzer} \citep{tappe06} confirm that line contributions at
24 $\mu$m are negligible. Our measured fluxes are presented in
Table~\ref{fluxtable}.

\section{Modeling}
\label{disc}

Because of the morphological similarities between the IR emission and
the blast wave seen in X-rays, we believe the dust present in the ISM
is being collisionally heated by electrons and protons in the outward
moving shock wave \citep{dwekarendt92}. The modeling of dust emission
for these remnants is similar to the modeling done on the four Type Ia
remnants in Paper I, where it is described more fully. Our model uses
a one-dimensional plane-shock approximation. The sputtering timescale,
$\tau_p$, is equal to $\int_0^t n_p dt$, and is one of the inputs to
the code. The other inputs are electron temperature $T_e$, ion
temperature $T_i$, gas density $n$, grain size distribution, and grain
composition and relative abundances. For the composition, abundances,
and distribution of dust grains, we follow WD, particularly the model consisting of
separate silicate and carbonaceous grain populations in fixed proportions,
 in a range of sizes from 1 nm to 1 $\mu$m.  We use the ``average'' LMC model with the
maximum amount of small carbonaceous grains. We do not model emission 
features from polycyclic aromatic  hydrocarbons (PAHs), as these emission 
features are not expected to contribute in MIPS bands. In order to test the dependency of our results on the 
dust grain-size distribution, we also consider an alternative model put forth 
by \citet{cartledge05}, hereafter C05. Dust masses derived from this
model were nearly identical to those from the WD model, but since the C05
model predicts a higher dust-to-gas
ratio in the LMC (0.45\%), this model actually exaggerates the observed dust deficit.
We thus report only the results from the WD model, and believe that these results are
conservative. X-ray analysis provides estimates of the electron temperature, ionization 
timescale $\tau_i \equiv \int_0^t n_e dt$, and emission measure $EM \equiv n_eM_g$
($M_g$ is mass of swept-up gas) of the plasma. We assume tight dust-gas coupling,
since the gyro-radii
of dust grains are small compared to the thickness of the shock \citep{dwek96}.
We used archival {\it Chandra} data for our analysis, and fit X-ray spectra using Sedov 
non-equilibrium ionization (NEI) thermal models in XSPEC \citep{arnaud96,borkowski01}.
Emission-measure averaged $T_e$ and $T_i$ from these models and the reduced $\tau_i$
($1/3$ of the Sedov $\tau_{ised}$, defined as the product of the postshock electron
density and the SNR age) were then used as inputs to our plane shock model.
(The approximate factor 1/3 arises from applying results of a spherical 
model to plane-shock calculations; see Fig.~4 of Borkowski et al.~2001).
Since the $T_i/T_e$ ratio is close to 1 in these models, we 
set $T_i = T_e$.

We employed two methods to model dust emission. In the first, we fix
$\tau_p$, $T_e$, and $T_i$ as derived from X-rays (taking $\tau_p =
\tau_i/1.2$), leaving only the density and total dust mass as free
parameters. In the second, we use the dynamical age of the remnant, as
derived from optical or global X-ray studies, leaving the shock age
and density as free, but correlated, parameters. We find that these
two different methods produce very similar results, within a factor of
$\sim 25\%$. We thus report only the inputs and results from the first
method. Model input parameters are given in Table~\ref{inputs}.  The
density of the gas is then adjusted to reproduce the observed 70/24 $\mu$m
flux ratio, and 24 $\mu$m flux is normalized to the observed value to
provide a total dust mass in the region of interest. The emission
measure divided by the electron density gives an estimate for the
amount of gas swept up by the blast wave, and dividing the dust mass
by the gas mass gives us a dust-to-gas mass ratio for the shocked
ISM. Finally, dust temperatures are calculated in our model. Because we
 use a range of grain sizes, we report a range of temperatures, since 
grains of different sizes will be heated differently under the same 
plasma conditions. The range of temperatures quoted is therefore the 
result of a single fit to the flux ratio. Results are summarized in 
Table~\ref{results}.

\subsection{N132D}

SNR N132D has been well studied in
optical wavelengths \citep{morse96}, and is one of the brightest
remnants in the Magellanic Clouds at X-ray wavelengths. 
The remnant is extraordinarily bright at 24 $\mu$m, where we find a total flux
of $\sim 3$ Jy, consistent with \citet{tappe06}. It is the only remnant in our sample that is brighter
at 24 $\mu$m than at 70 $\mu$m. This implies warm dust, and therefore a dense
environment (to provide the inferred heating), consistent with the remnant being very
bright in X-rays. \citet{morse96} estimate a preshock hydrogen density for this 
remnant of 3 cm$^{-3}$ based on modeling of the photoionized shock 
precursor. 
We analyzed the NW rim and the bright southern rim separately. We find high
densities, in good agreement with the expected postshock proton 
density $n_p$ of 12 cm$^{-3}$. Densities are higher by a factor of $\sim 2$ in
the NW, and comparatively less mass in gas in that region. The mass in dust, however, was
comparable to what was found in the south, adjusting for the different
sizes of the regions.

\subsection{N49B}

N49B is believed to be older than N132D, perhaps
as old as 10,000 yrs.  
The ionization age is lower than this, however,
perhaps due to an explosion into a pre-existing cavity
\citep{hughes98}. This remnant is much fainter than N132D at all
wavelengths, but especially so at 24 $\mu$m, and the flux ratio is lower
 by more than order of magnitude in N49B compared to N132D. This can 
be explained by the much lower density we find in N49B: $n_{p}$ = 1.1 cm$^{-3}$ 
from our dust models, and $n_{p}$ = 2.1 cm$^{-3}$ from a Sedov analysis of X-ray
data. Since dust heating rates are strongly dependent on density, a
low-density environment will result in much cooler dust, whose
spectrum peaks at longer wavelengths. Despite the contrasts in
density and flux ratio of more than an order of magnitude, the
dust-to-gas ratio in this remnant is only two times lower than that found in
N132D.

\subsection{N23}

At 70 $\mu$m, the brightest parts of the shell of N23
are clearly visible. The region selected for analysis completely
enclosed the emission visible at 70 $\mu$m, which contained about 75\%
of the emission visible at 24 $\mu$m. Dust modeling yields $n_{p}$ = 5.8 cm$^{-3}$.
\citet{hughes06} found densities of 10 and 23 cm$^{-3}$ in a couple of X-ray bright 
shocks in this region of the remnant.

\subsection{0453--68.5}

Although 0453--68.5 had the highest 70/24 $\mu$m ratio of any
remnant in this sample, only the north rim of the shell was clearly
separable from the background at 70 $\mu$m. This is due to the overall
faintness of the remnant, the high levels of background in
the region, and the lower sensitivity of the 70 $\mu$m
array. Nonetheless, there was sufficient S/N to analyze the north rim
of the remnant. The density found from modeling dust emission, 
$n_{p}$ = 0.63 cm$^{-3}$,
 was also the lowest of these remnants. Sedov modeling of X-rays yielded a 
value of $n_{p}$ = 1.1 cm$^{-3}$. Both the temperature and density are
 consistent with the observations of overall faintness at 24 $\mu$m and a high 
70/24 $\mu$m ratio.

\section{Discussion and Conclusions}
\label{concls}

We find excellent (generally within a factor of 2) agreement between 
densities derived from dust modeling and estimated from optical and X-ray 
data. As has been proposed in the past \citep[e.g.,][]{dwekarendt92}, dust 
emission is indeed a valuable density diagnostic of X-ray emitting plasmas.
The good match between X-ray and IR derived densities suggests that the
X-ray emitting gas is not highly clumped.

Our basic quantitative results are contained in Table~\ref{results}.
First, note that the total mass in dust we infer is quite small, of
order 0.01--0.1 $\msun$, for all objects.  Morphological
evidence suggests that this dust is associated with the blast wave, but
these values also serve as restrictive upper limits for the amount of
ejecta dust that can have been produced in these core-collapse
supernovae. They are, for the most part, less than the $0.08 \msun \lesssim M_{d} 
\lesssim 0.3 \msun$ calculated by \citet{todini01}. Substantial amounts of 
dust can be destroyed when the ejecta are reverse-shocked, but the ejecta dust
mass ultimately delivered to the ISM appears to be quite small. This is 
consistent with the findings of \citet{stanimirovic05}, who analyzed
IR emission from the LMC SNR 1E 0102.2-7219.

We find that on average, about 40\% of the mass in dust grains,
including $\sim 90\%$ of the mass in grains smaller than 0.04 $\mu$m,
has been destroyed via sputtering in these remnants.  After accounting
for this sputtering, we find an initial dust-to-gas ratio that is
lower by a factor of roughly four than what is generally expected for the
LMC. The values in Table~\ref{results} are similar to the dust-to-gas ratios we 
found in remnants from Type Ia SNe, and they are still well below the value 
reported in WD of $\sim 2.5 \times 10^{-3}$. In Paper I we speculated that low 
dust/gas ratios might be expected for Type Ia remnants expanding into 
lower-than-average density material, in which previous supernovae might have
destroyed some grains. Since we now have a similar deficit for core-collapse
remnants expanding into denser media, this explanation is less likely, although
it is still possible that the dust content is generally low in the vicinity
of core-collapse SNe.

The overall lower dust/gas ratios we infer will require
another explanation. It is possible that sputtering timescales or rates
might have been underestimated. Our models
indicate that sputtering timescales would have to be increased by an
order of magnitude, on average, to match the expected dust-to-gas
ratio in the WD model, an unlikely possibility.  Various factors
could increase sputtering rates. Sputtering can be enhanced by increased 
relative velocities between ions and grains, either caused by grain 
motions (not included in our models) or by higher than assumed ion 
temperatures. In order to estimate the importance of such effects, we doubled ion 
temperatures in our shock models, resulting in only 10--15\%\ increase in the fraction 
of the original dust destroyed for the relatively fast shocks considered here. Sputtering
is expected to be enhanced for nonspherical grains with increased
surface-to-volume ratio $S/V$.  For a prolate ellipsoidal grain with
an axial ratio of 2, $S/V$ is twice as large as for a spherical grain,
which enhances sputtering rates by a modest factor of 2. We find it unlikely that these modest
enhancements in sputtering rates can account for the lower than
expected dust-to-gas mass ratios.

%It is also possible that the dust mass in the WD dust model (and more
%generally in most dust models considered in the past) may be
%overestimated. The total grain volume per H atom in the Milky Way
%deduced from depletion of elements onto dust in the ISM
%\citep[assuming solar abundances;][]{jenkins04} is only $\sim 60\%$ of
%the total grain volume in the WD model \citep{draine04}. This
%discrepancy becomes less severe if oversolar abundances are assumed
%for the ISM. For example, \citet{cartledge06} favor oversolar
%abundances based on observations of young F and G stars and the
%observed depletion patterns.

Porous (fluffy) grains provide a possible explanation for the low 
dust-to-gas ratios. Porous grains \citep{mathis96} can be characterized by a porosity
(fraction of grain volume devoid of material) $\cal P$.  Such grains have an
extinction per unit mass generally larger than for spherical compact
grains. \citet{vosh06} modeled visual and UV absorption towards
$\zeta$ Oph and $\sigma$ Sco with a mixture of highly porous 
(${\cal P} \gapprox 0.9$) and compact grains with dust mass of only 70\%\ and
44\%\ of the dust mass in the WD model.

We expect enhanced sputtering rates for porous grains compared
with spherical compact grains. A highly porous (${\cal P}=0.9$) grain
has surface area 4.6 times larger than a compact grain of the same
mass, so for the same sputtering yields an enhancement of sputtering rates by a
factor of 4.6 is expected. In addition, protons, $\alpha$ particles, and heavier
ions can penetrate much deeper into porous grains, because their
range within the solid is inversely proportional to the mean grain
density. A proton with energy $E$ will penetrate $0.09 (E/1~{\rm
keV})$ $\mu$m into a porous silicate grain with ${\cal P}=0.9$ (we
used eq.~18 of Draine \&\ Salpeter 1979 to estimate the proton
projected range $R_p$). \citet{jurac98} found that sputtering yields
are enhanced if the grain radius $a$ is less than $3R_p$ because of
additional sputtering from the back and sides of the grain. Grains
as large as $0.25 (E/1~{\rm keV})$ $\mu$m are then
destroyed more efficiently.  It is also possible that the rough
surface geometry expected for porous grains could result in a further
enhancement of sputtering rates, as the protruding grain extensions
comparable in size to the proton or $\alpha$ particle projected ranges
could be more readily sputtered \citep{jurac98}. An order of magnitude
enhancement in sputtering rates is thus possible for highly porous grains.

The reduced preshock dust content and substantial enhancement in
sputtering rates that result from porous grain models provide a
promising explanation for the apparent dust deficit in LMC SNRs. These
effects might be partially offset by the increased IR emissivity of porous
grains \citep{vosh06}, as less dust is then required to account for the observed IR
emission. It is possible, however, that 
compact grains dominate the observed IR emission, if porous grains (or 
porous mantles surrounding compact cores) have been preferentially destroyed.
Modeling of {\it Spitzer} observations
with porous grain models is required to assess whether these models can resolve the
apparent dust deficit in LMC SNRs, including SNR 1987A. Spectroscopic follow-up
on these remnants in the far-IR is required to confirm and strengthen
our results. Implications of a resolution go beyond these particular SNRs or
even the LMC as a whole and could help bring about a significant
improvement in our understanding of interstellar dust in general.

\acknowledgments

This work was supported by NASA through Spitzer Guest Observer grant
RSA 1265236. We thank an anonymous referee for helpful comments.

\clearpage

\begin{deluxetable}{lccc}
%\vspace{-0.3truein}

\tablecolumns{4}
\tablewidth{0pc}
\tabletypesize{\footnotesize}
\tablecaption{Measured Fluxes}
\tablehead{
\colhead{Object} & 24 $\mu$m & 70 $\mu$m & 70/24}

\startdata
N132D NW   & 730 $\pm${73} & 430 $\pm${96} & 0.59 $\pm${0.13}\\
N132 S     & 1000 $\pm${100} & 770 $\pm${170} & 0.76 $\pm${0.17}\\
N49B       & 43 $\pm${4.3} & 395 $\pm${79} & 9.1 $\pm${2.0}\\
0453--68.5 & 13 $\pm${1.3} & 250 $\pm${50} & 19 $\pm${4.2}\\
N23        & 100 $\pm${10} & 240 $\pm${50} & 2.4 $\pm${0.5}\\

\enddata

\tablenotetext{a}{All fluxes given in millijanskys}

\label{fluxtable}
\end{deluxetable}

\begin{deluxetable}{lccc}
%\vspace{-0.2truein}
\tablecolumns{6}
\tablewidth{0pc}
\tabletypesize{\footnotesize}
\tablecaption{Model Inputs}
\tablehead{
\colhead{Object} & $T_e(=T_i)$  & $\tau_{p}$ & Age \\
& (keV) & ($10^{3}$ cm$^{-3}$ yr) &(yr) 
} 

\startdata
N132D NW   & 0.61 & 17 & 2500{\tablenotemark{a}}\\
N132D S    & 1.02 & 3.2 & 2500\\
N49B       & 0.36 & 2.4 & 10,900{\tablenotemark{b}}\\
0453--68.5 & 0.29 & 8.6 & 8700{\tablenotemark{b}}\\
N23        & 0.56 & 6.7 & 4600{\tablenotemark{c}}\\

\enddata

\tablenotetext{a}{\citet{morse96}}
\tablenotetext{b}{\citet{gaensler03}}
\tablenotetext{c}{\citet{hughes06}}

\label{inputs}
\end{deluxetable}

\begin{deluxetable}{lcccccccc}
%\vspace{-0.2truein}
\tablecolumns{7}
\tablewidth{0pc}
\tabletypesize{\footnotesize}
\tablecaption{Model Results}
\tablehead{
\colhead{Object} & $n_{p}$ (cm$^{-3})$ & T(dust) (K) & Dust Mass ($M_\odot$) & $n_eM_g$ ($cm^{-3}M_\odot$) & Dust/gas & \% dest. & Dust/Gas (orig.) & $L_{36}$}

\startdata
N132D NW & 34 & 95-120 & 0.0075 & 660 & $4.6 \times 10^{-4}$ & 50 & $9.2 \times 10^{-4}$ & 35\\
N132D S & 14 & 85-105 & 0.015 & 1020 & $2.5 \times 10^{-4}$ & 38 & $4.0 \times 10^{-4}$ & 94\\
N49B & 1.1 & 50-60 & 0.08 & 580 & $1.8 \times 10^{-4}$ & 27 & $2.4 \times 10^{-4}$ & 8.4\\
0453--68.5 & 0.63 & 40-55 & 0.10 & 120 & $6.1 \times 10^{-4}$ & 33 & $9.8 \times 10^{-4}$ & 4.5\\
N23 & 5.8 & 70-85 & 0.011 & 250 & $3.0 \times 10^{-4}$ & 39 & $4.9 \times 10^{-4}$ & 8.7\\

\enddata

\tablecomments{Col (2): postshock proton density; col. (3): for 0.02-0.1 $\mu$m grains; 
col (4): mass of dust currently observed, after sputtering (post-shock), in $\msun$;
 col (6): ratio of current dust mass to swept-up gas mass; col (7):
percentage of dust destroyed via sputtering; col (8): inferred dust-to-gas mass ratio in preshock ISM;
col (9): $L_{36} \equiv L_{IR}/10^{36}$ erg s$^{-1}$.} 

\label{results}
\end{deluxetable}

\begin{figure}
\plotone{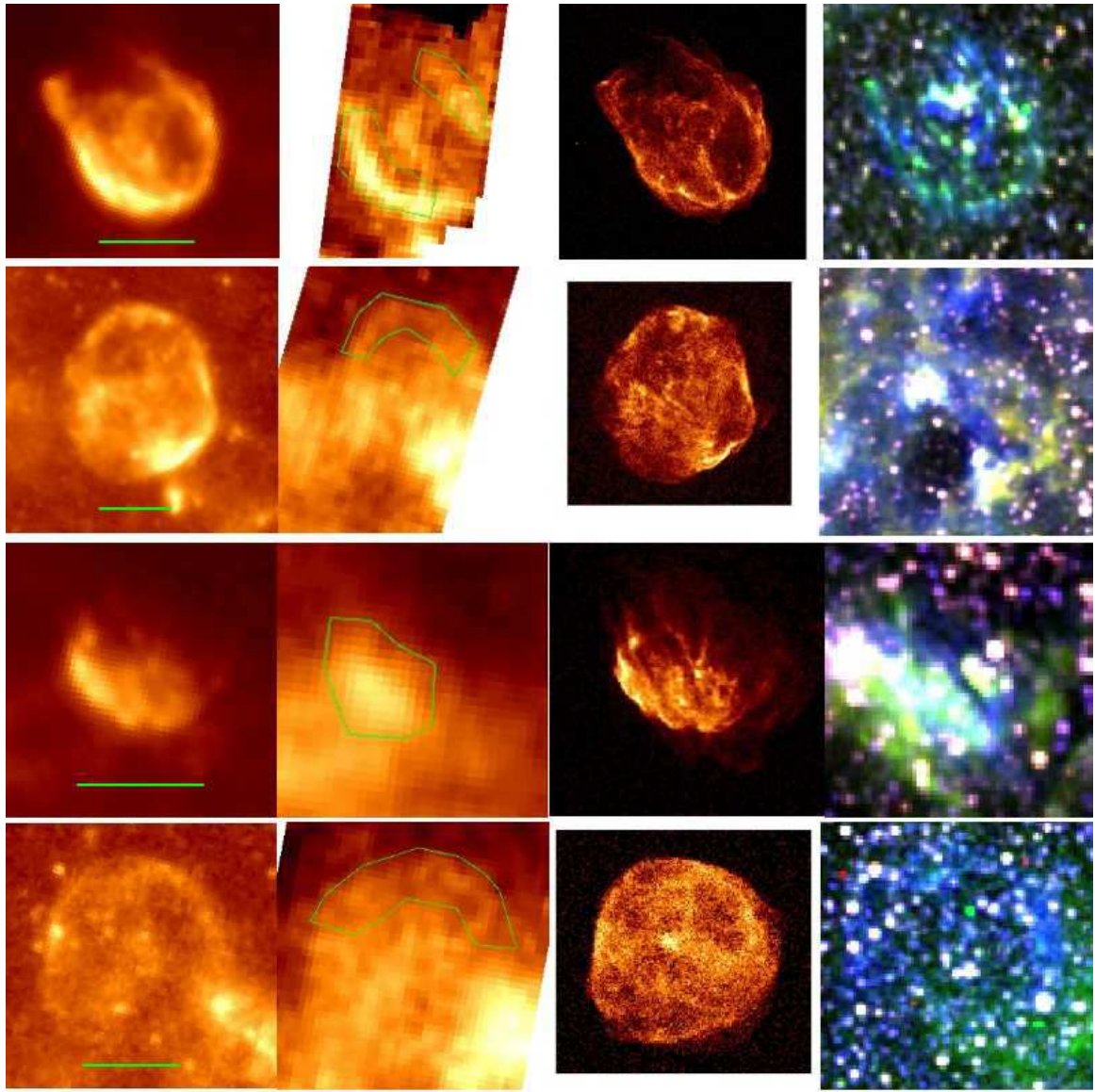}
\caption{Top row, from left to right: N132D at 24 and 70 $\mu$m (the region of
interest at 70 microns is marked on the image), in the X-rays (broadband,
Chandra image) and in the optical (overlay of MCELS images, with [S II],
H$\alpha$ and [O III] marked in red, green and blue, respectively).   Second,
third and fourth rows show the same sequence for N49B, N23 and 0453-68.5,
respectively. One arc-minute scales are shown on the 24 $\mu$m images.
\label{images}
}
\end{figure}

\end{document}